\documentclass[letterpaper]{article} 
\usepackage{aaai2027}  
\nocopyright

\usepackage[hyphens]{url}  
\usepackage{graphicx} 
\usepackage{natbib}  
\usepackage{caption} 
\usepackage{algorithm}
\usepackage[noend]{algorithmic}
\usepackage{braket}
\usepackage{newfloat}
\usepackage{listings}
\DeclareCaptionStyle{ruled}{labelfont=normalfont,labelsep=colon,strut=off} 
\floatstyle{ruled}
\newfloat{listing}{tb}{lst}{}
\floatname{listing}{Listing}

\usepackage{booktabs}

\usepackage{multirow}
\usepackage{makecell}

\usepackage{amsmath}
\newcommand{\argmin}{\mathop{\rm arg~min}\limits}

\title{Optimizing Locally Equivalent Graph States in Quantum Computing \\ Using Greedy Best-First Search}
\author {
    Sora Harada\textsuperscript{\rm 1,\rm 2},
    Ryo Kuroiwa\textsuperscript{\rm 2, \rm 1},
    Shin Nishio\textsuperscript{\rm 3, \rm 4},
    Takeaki Uno\textsuperscript{\rm 2, \rm 1}\corresponding
}
\affiliations {
    \textsuperscript{\rm 1}The Graduate University for Advanced Studies, SOKENDAI, 
    Tokyo, Japan
    
    \textsuperscript{\rm 2}National Institute of Informatics,
    Tokyo, Japan
    
    \textsuperscript{\rm 3}University College London,
    London, United Kingdom

    \textsuperscript{\rm 4}Keio University,
    Yokohama, Japan

    sora@nii.ac.jp, kuroiwa@nii.ac.jp, parton@sfc.wide.ad.jp, uno@nii.ac.jp
}

\begin{document}
\pagestyle{plain}

\maketitle

\begin{abstract}
    In measurement-based quantum computation and quantum repeater networks,
    graph-structured quantum states called graph states play a central role as an essential resource.
    In a graph state, each vertex corresponds to a qubit, and each edge is prepared by applying one controlled-Z (CZ) gate.
    Local complementation, which toggles adjacencies among the neighbors of a selected vertex, produces operationally equivalent graph states up to an inexpensive operation called local Clifford gates. Therefore, finding equivalent graphs up to local complementation with fewer edges and lower maximum degree can reduce the preparation cost, such as the number of CZ gates and the circuit depth.
    In this paper, we formulate the problem of finding a desirable equivalent graph via local complementations as a state space search problem.
    We apply a heuristic search algorithm, greedy best-first search (GBFS), to the formulated problem.
    Our experimental evaluation shows that despite its simplicity, GBFS outperforms existing approaches based on integer linear programming or local search across multiple graph classes, including random graphs, bounded-degree graphs, and most of the graphs constructed from quantum LDPC codes.
    In addition, we empirically analyze GBFS's search behavior, confirming that we need search rather than stopping at the first local minimum to find the best solution.
\end{abstract}

\section{Introduction}
\label{sec:intro}
Quantum computing has attracted significant attention for its potential to provide computational speedups from the perspective of computational complexity theory~\cite{shor1994algorithms, grover1996fast} by exploiting quantum mechanical phenomena such as superposition and entanglement.
\emph{Graph states}
provide a natural and mathematically tractable framework for describing entanglement resources as undirected graphs \cite{hein2004multiparty}.
In measurement-based quantum computation, graph states serve as computational resources, and arbitrary quantum computation can be translated into graph states with suitable measurement instructions~\cite{raussendorf2003measurement, browne2007generalized}. Hence, optimizing graph states corresponds to arbitrary quantum circuit optimization.
Graph states are also essential resource states to distribute quantum entanglement over long distances without requiring long-lived quantum memories~\cite{azuma2015all}.

\begin{figure}
    \centering
    \includegraphics[width=0.7\linewidth]{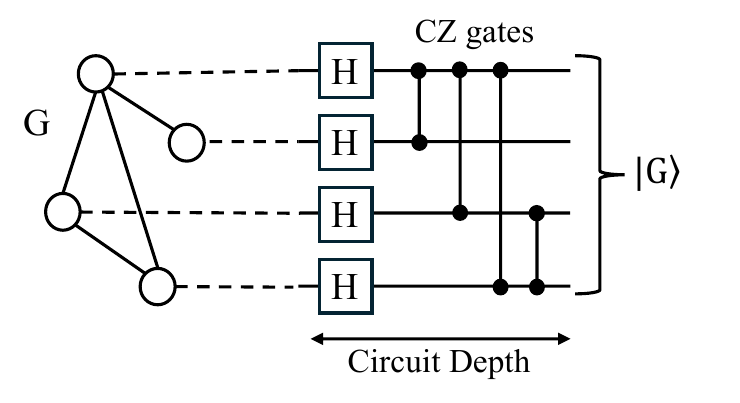}
    \caption{Quantum circuit for the preparation of a graph state $\ket{G}$ for graph $G$. Each horizontal line corresponds to a qubit prepared in $\ket{0}$ state.
    For preparing a graph state, one needs to apply Hadamard gates (H) in parallel first, and then apply CZ gates corresponding to the edges.}
    \label{fig:preparation_circuit}
\end{figure}

As shown in our example (Fig.~\ref{fig:preparation_circuit}), given a graph, its corresponding graph state is prepared by applying one controlled-Z (CZ) gate for each edge on qubits corresponding to the vertices.
The number of edges directly gives the total number of two-qubit gates required.
The maximum degree is related to the circuit depth; a qubit corresponding to a vertex of degree $d$ must participate in $d$ CZ gates, and these gates cannot be applied simultaneously to the same qubit. Therefore, the CZ depth is at least the maximum degree. Since CZ gates on disjoint edges can be executed in parallel, the exact minimum depth $D(G)$ of graph $G$ satisfies
$d_{\rm max}(G) \leq D(G) \leq d_{\rm max}(G)+1$, where $d_{\rm max}(G)$ is the maximum degree of $G$.

The number of CZ gates to prepare a graph state can be reduced by applying quantum operations.
In particular, \emph{local complementation} of the graph, which toggles adjacencies among the neighbors of a selected vertex, can possibly reduce the number of edges. Local complementation can be implemented via \emph{local Clifford} operations on graph states.
Since local Clifford operations are typically cheaper than entangling operations such as the CZ gate, reducing the number of edges and the maximum degree using a sequence of local complementations has recently been studied in the quantum computing literature \cite{Sharma2026,connolly2026efficient}.

Reducing these CZ operations is desirable both for near-term quantum devices~\cite{preskill2018quantum}, where two-qubit gate errors are a major concern, and for fault-tolerant quantum computing~\cite{gottesman1997stabilizer}, where logical entangling operations require physical resources. For example, in surface-code-based fault-tolerant quantum computing architectures, logical entangling gates can be implemented through lattice surgery~\cite{horsman2012surface}, which incurs space and time overhead for joint measurements and auxiliary qubits~\cite{litinski2019game}. Reducing the number of entangling gates can therefore help reduce the space-time resource requirements of graph-state preparation.

In this paper, we formulate the problem of transforming a graph to one with fewer edges and a smaller maximum degree using local complementations as a state space search problem \cite{Russel2020}.
Then, we apply greedy best-first search (GBFS) \cite{Doran1966} to the problem, using graph isomorphisms for symmetry breaking during search.
While GBFS is a satisficing search method in conventional state space search, it works as an anytime algorithm with a guarantee of optimality in our problem setting.
We conduct a comprehensive empirical evaluation, using the graph classes from the previous work and practical instances derived from datasets of quantum error-correcting codes.
We demonstrate that despite its simplicity, GBFS outperforms existing approaches based on integer linear programming or local search.
In addition, we show that while GBFS succeeds in finding an optimal solution much earlier than exhausting the search space, it still explores multiple local minima; local search stopping at the first local minimum is not sufficient.

\section{Related Work}
\label{sec:related}

The problem of reducing the number of edges in a graph state has been studied in previous work.
Although whether a graph can be transformed into another graph using a sequence of local complementations can be decided in polynomial time \cite{bouchet1993recognizing,van2004graphical}, it is not known whether a polynomial-time algorithm can minimize the number of edges.
\citet{adcock2020mapping} enumerated graphs that can be obtained by a sequence of local complementations starting from every graph with up to 12 vertices.
\citet{Sharma2026} proposed an exact integer linear programming (ILP) model to minimize the number of edges using local complementations.
They also developed a method based on simulated annealing \cite{Kirkpatrick1983} and used it to warm-start the ILP model.
\citet{connolly2026efficient} proposed a local search algorithm using local complementations.
In addition, they developed a graph decomposition method using quotient-augmented strong split tree (QASST).

Different but related problems on graph states were also studied.
\citeauthor{duncan2020graph}~\shortcite{duncan2020graph} utilized ZX-calculus as an intermediate representation for the quantum circuit and applied local complement-based rules to simplify graph states. 
\citet{doherty2026fast} used heuristic search algorithms, Monte-Carlo tree search (MCTS) and beam search,\footnote{They also used ``(greedy) best-first search'', but it is a greedy algorithm without backtracking and differs from the conventional definition of best-first search in the heuristic search literature.} to prepare a graph state with CX, CY, and CZ gates.
In particular, MCTS was combined with reinforcement learning to guide the search.
\citet{doherty2026fast} also used the simulated annealing method by \citet{Sharma2026} as preprocessing to reduce the number of edges in the target graph state.

Previous work investigated applications of heuristic search in quantum computing, in addition to \citet{doherty2026fast}.
\citet{Liu2026} applied MCTS, combined with reinforcement learning, to quantum error correction.
Similarly, combinations of heuristic search and reinforcement learning were applied to quantum circuit synthesis \cite{Chen2024,Bao2024,Turner2025}.
In addition, heuristic search was applied to quantum circuit compilation \cite{Zulehner2019, nishio2020extracting}.
Automated planning \cite{Ghallab2016}, a technology closely related to heuristic search, has also been applied to quantum circuit synthesis \cite{Shaik2026} and quantum circuit compilation \cite{Venturelli2017,Booth2018,Do2020,Shaik2023}.
However, to our knowledge, previous work has not applied heuristic search to optimization of graph states using only local complementations.

\section{Problem Definition}
\label{sec:problem}
\begin{figure}[t]
    \centering
    \includegraphics[width=0.4\linewidth]{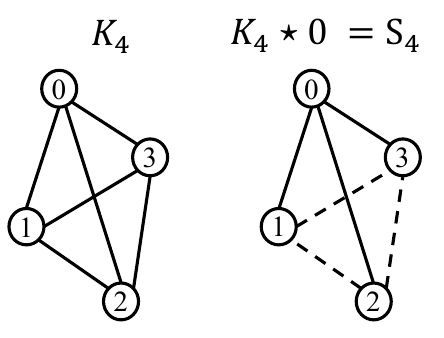}
    \caption{Local complementation on vertex $0$ in a complete graph with 4 vertices ($K_4$) yields a star graph ($S_4$).}
    \label{fig:local_complement}
\end{figure}

Let $G$ be an undirected graph, with the set of vertices $V(G)$ and the set of undirected edges $E(G) \subseteq V(G) \times V(G)$.
The \emph{local complementation} with regard to $v \in V(G)$ transforms $G$ into another graph $G \star v$, where the adjacencies among the neighbors of $v$ are toggled, as shown in Fig.~\ref{fig:local_complement}.

\begin{figure}[t]
    \centering
    \includegraphics[width=0.8\linewidth]{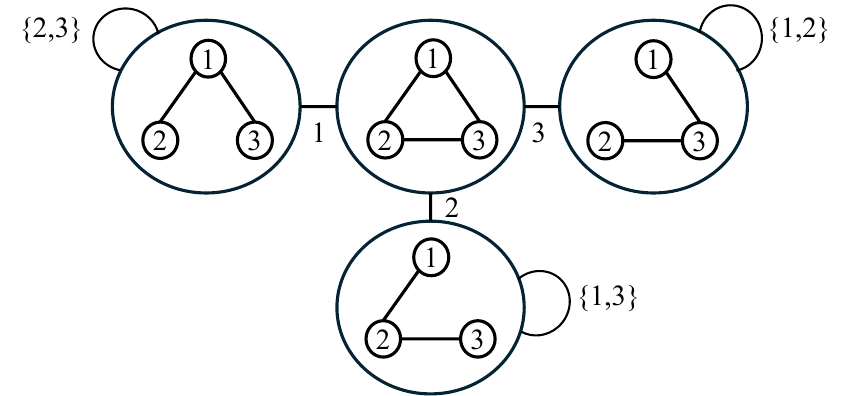}
    \caption{LC orbit for a complete graph with 3 vertices ($K_3$). Each graph is connected via a local complementation. LC orbit may contain self-loops.}
    \label{fig:lcorbit}
\end{figure}

Formally, the set of neighbors of $v \in V(G)$ is defined as $N_G(v) = \{ u \in V(G) \mid \{ u, v \} \in E(G) \}$.
The graph $G \star v$ has the set of vertices $V(G \star v) = V$ and the set of edges 
\begin{align*}
    & E(G \star v) = \\
    & (E(G) \setminus  \{ \{ u, x \} \in E(G) \mid u, x \in N_G(v), u \neq v \}) \\
    & \cup \{ \{ u, x \} \not\in E(G) \mid u, x \in N_G(v), u \neq v \}.
\end{align*}
Two graphs are \emph{locally equivalent} if one can be obtained by a sequence of local complementations from another.
The \emph{local-Clifford (LC) orbit} of $G$ is the set of graphs that are locally equivalent to $G$, including $G$ \cite{adcock2020mapping}.
We visualize an example LC orbit in Fig.~\ref{fig:lcorbit}.
In our setting, given a graph $G$, we want to find a sequence of local complementations that transforms $G$ into a locally equivalent graph with fewer edges.
In addition, we also want to reduce the maximum degree, defined as $d_\text{max}(G) = \max_{v \in V(G)} | \{ \{ u, v \} \in E(G) \} |$ for a graph $G$, as a secondary objective.

In terms of computational complexity, the weighted version of the edge minimization problem, which is to find a locally equivalent graph minimizing the total edge weight, is proved to be NP-hard, and its decision version is NP-complete \cite{Sharma2026}.
However, the complexity of the unweighted version, where we minimize the number of edges, remains an open question; the proof for the weighted version does not straightforwardly transfer to the unweighted version, and proving it is out of scope of this paper.

\section{Heuristic Search in the LC Orbit}

We consider the problem of finding a locally equivalent graph with fewer edges (and a smaller maximum degree) as a state space search problem \cite{Russel2020}, where a state is a graph and an action is a local complementation; the search space is the LC orbit.
In typical state space search problems, such as puzzle-solving and planning, the objective is to find a sequence of actions that transforms the initial state into a goal state, and an optimization metric is defined with respect to the action sequence, e.g., minimizing the number of actions or the total action cost.
Although a sequence of local complementations induces a sequence of local Clifford operations, these operations can be composed into a single local Clifford operation on each qubit~\cite{hein2004multiparty, van2004graphical}. Therefore, the number of local complementations does not directly determine the cost of implementing the resulting graph state, and we do not optimize the sequence length.
In our problem, any locally equivalent graph with fewer edges can be treated as the final state, and the optimization metric depends only on it.

We adopt greedy best-first search (GBFS) \cite{Doran1966}, a heuristic search algorithm that prioritizes states to search using a heuristic function $h$.
In particular, for graph $G$, we use $h(G) = (|E(G)|, d_\text{max}(G))$ as a heuristic value and use the lexicographic order.
Given two graphs $G$ and $G'$, $h(G) < h(G')$ iff $|E(G)| < |E(G')|$ or $|E(G)| = |E(G')| \land d_\text{max}(G) < d_\text{max}(G')$.

\begin{algorithm}[t]
    \begin{algorithmic}[1]
        \STATE $\pi(G_\text{init}) \leftarrow \langle \rangle$, $h(G_\text{init}) \leftarrow (|E(G_\text{init})|, d_\text{max}(G_\text{init}))$ \label{alg:gbfs:pi-init}
        \STATE $O \leftarrow \{ G_\text{init} \}, C \leftarrow \{ \}, G_\text{best} \leftarrow G_\text{init}$ \label{alg:gbfs:init-list}
        \WHILE{$O \neq \emptyset$ and the resource limit is not reached} \label{alg:gbfs:check-open}
            \STATE Let $G \in \argmin_{G' \in O} h(G')$ \label{alg:gbfs:remove}
            \STATE $O \leftarrow O \setminus \{ G \}, C \leftarrow C \cup \{ G \}$ \label{alg:gbfs:move}
            \FORALL{$v \in V(G)$} \label{alg:gbfs:expand}
                \IF{$G \star v \not\in O \cup C$} \label{alg:gbfs:check-lists}
                    \STATE $\pi(G \star v) \leftarrow \pi(G)$ extended with $v$ \label{alg:gbfs:pi}
                    \STATE $h(G \star v) \leftarrow (|E(G \star v)|, d_\text{max}(G \star v))$ \label{alg:gbfs:h}
                    \IF{$h(G \star v) < h(G_\text{best})$} \label{alg:gbfs:check-best}
                        \STATE $G_\text{best} \leftarrow G \star v$ \label{alg:gbfs:update-best}
                    \ENDIF
                    \STATE $O \leftarrow O \cup \{ G \star v \}$ \label{alg:gbfs:push}
                \ENDIF
            \ENDFOR
        \ENDWHILE
        \RETURN $\pi(G_\text{best})$ \label{alg:gbfs:regturn}
    \end{algorithmic}
    \caption{GBFS in the LC Orbit}
    \label{alg:gbfs}
\end{algorithm}

Algorithm~\ref{alg:gbfs} shows pseudo-code of our GBFS.
Given a graph $G_\text{init}$, we maintain $\pi(G)$, a sequence of local complementations transforming $G_\text{init}$ to a locally equivalent graph $G$.
We initialize $\pi(G_\text{init})$ to an empty sequence (line~\ref{alg:gbfs:pi-init}).
We put the states to search in the \emph{open list} $O$, initialized with $\{ G_\text{init} \}$, and the searched states in the \emph{closed list} $C$, initialized with an empty set (line~\ref{alg:gbfs:init-list}).
We also maintain the best state $G_\text{best}$, initialized with $G_\text{init}$.
In each iteration, we move a state $G$ minimizing $h(G)$ from the open list to the closed list (lines~\ref{alg:gbfs:remove} and \ref{alg:gbfs:move}). 
Then, we \emph{expand} the state $G$, i.e.,  generate a \emph{successor state} $G \star v$ for each vertex $v \in V(G)$ (line~\ref{alg:gbfs:expand}).
If the successor state is not in the open or closed lists, then it has not been seen yet.
We extend $\pi(G)$ to $\pi(G \star v)$, compute the heuristic value $h(G \star v)$, update the best state $G_\text{best}$ if $G \star v$ is better, and add $G \star v$ into the open list (lines~\ref{alg:gbfs:check-lists}--\ref{alg:gbfs:push}).
We repeat the above procedure until the open list is exhausted or the resource limit (e.g., time or memory) is reached (line~\ref{alg:gbfs:check-open}).
We finally return $\pi(G_\text{best})$, the sequence of local complementations to reach $G_\text{best}$ (line~\ref{alg:gbfs:regturn}).
Since every graph in the LC orbit is reachable from $G_\text{init}$ and the LC orbit is finite, GBFS, which detects duplicate states using the open and closed lists, generates every graph in the LC orbit before the open list becomes empty. Therefore,
if the open list is empty at termination, $G_\text{best}$ lexicographically minimizes the number of edges and the maximum degree.

\subsection{Symmetry Breaking via Graph Isomorphisms}

In our implementation, we represent a state as an adjacency matrix implemented by a bit vector and use a hash table to detect duplicate states in the open and closed lists 
(lines~\ref{alg:gbfs:move}, \ref{alg:gbfs:check-lists}, and \ref{alg:gbfs:push} in Algorithm~\ref{alg:gbfs}).
In practice, two graphs with different adjacency matrices can be equivalent if they are symmetric; they actually have the same structure but with different vertex indices.
Formally, graphs $G$ and $H$ are \emph{isomorphic} if there exists an \emph{isomorphism}, a bijection $f : V(G) \to V(H)$ mapping a vertex $v \in V(G)$ to $f(v) \in V(H)$ such that $f(u)$ and $f(v)$ are adjacent ($\{ f(u), f(v) \} \in E(H)$) if and only if $u$ and $v$ are adjacent ($\{ u, v \} \in E(G)$).
If two graphs are isomorphic, we only need to consider one of them to find a locally equivalent graph that minimizes the number of edges and the maximum degree.

To detect isomorphic graphs, we use graph canonization, which transforms all isomorphic graphs into the same representation called the canonical form.
We store the canonical forms in the hash table for duplicate detection.
This approach can be viewed as an adaptation of symmetry breaking in heuristic search for automated planning \cite{Pochter2011,Domshlak2012}.Unlike these methods, which keep each state together with its canonical form, we store only the canonical forms and expand the canonical forms themselves, which reduces memory consumption as in \emph{orbit search}~\cite{Domshlak2015symmetry}.
While duplicate detection via graph isomorphisms can reduce the number of states, graph canonization may introduce computational overhead.
In our experiment, we evaluate GBFS with and without this mechanism.

\section{Empirical Evaluation}
\begin{table*}[htb]
    \centering
    \small
    \setlength{\tabcolsep}{1mm}
    \begin{tabular}{lccccccccccccccccccccc}
    \toprule
    && \multicolumn{2}{c}{LS} & \multicolumn{3}{c}{LSSF} & \multicolumn{4}{c}{SA-ILP} & \multicolumn{5}{c}{GBFS} & \multicolumn{5}{c}{Orbit} \\
    \cmidrule(lr){3-4} \cmidrule(lr){5-7} \cmidrule(lr){8-11} \cmidrule(lr){12-16} \cmidrule(lr){17-21}
    p & $|V|$ & $|E|$ & $d_{\max}$ & $|E|$ & $d_{\max}$ & \#TL & $|E|$ & $d_{\max}$ & \#Opt. & \#TL & $|E|$ & $d_{\max}$ & \#Opt. & \#TL & \#OM & $|E|$ & $d_{\max}$ & \#Opt. & \#TL & \#OM \\
    \midrule
    0.2 & 14 & 0.95 & 0.91 & 1.21 & 0.89 & 0 & \textbf{0.90} & 0.85 & 96 & 4 & \textbf{0.90} & \textbf{0.81} & \textbf{100} & 0 & 0 & \textbf{0.90} & \textbf{0.81} & \textbf{100} & 0 & 0 \\
    & 16 & 0.96 & 0.94 & 1.15 & 0.93 & 0 & \textbf{0.92} & 0.88 & 75 & 25 & \textbf{0.92} & \textbf{0.85} & \textbf{100} & 0 & 0 & \textbf{0.92} & \textbf{0.85} & 85 & 15 & 0 \\
    & 18 & 0.96 & 0.93 & 1.08 & 0.92 & 0 & \textbf{0.93} & 0.89 & 34 & 66 & \textbf{0.93} & \textbf{0.86} & \textbf{57} & 17 & 26 & \textbf{0.93} & \textbf{0.86} & 8 & 92 & 0 \\
    & 20 & 0.97 & 0.95 & 1.04 & 0.94 & 0 & 0.94 & 0.90 & \textbf{7} & 93 & \textbf{0.93} & \textbf{0.86} & 2 & 1 & 97 & \textbf{0.93} & \textbf{0.86} & 0 & 100 & 0 \\
    & 30 & 0.99 & 0.99 & 1.00 & 1.00 & 100 & \textbf{0.98} & 0.98 & 0 & 100 & \textbf{0.98} & \textbf{0.95} & 0 & 0 & 100 & \textbf{0.98} & \textbf{0.95} & 0 & 26 & 74 \\
    & 40 & \textbf{1.00} & 1.00 & \textbf{1.00} & 1.00 & 100 & \textbf{1.00} & \textbf{0.99} & 0 & 100 & \textbf{1.00} & \textbf{0.99} & 0 & 0 & 100 & \textbf{1.00} & \textbf{0.99} & 0 & 0 & 100 \\
    \midrule        
    0.5 & 14 & 0.80 & 0.77 & 0.80 & 0.77 & 0 & 0.65 & 0.63 & 0 & 100 & \textbf{0.62} & \textbf{0.57} & \textbf{100} & 0 & 0 & \textbf{0.62} & \textbf{0.57} & \textbf{100} & 0 & 0 \\
    & 16 & 0.80 & 0.79 & 0.80 & 0.79 & 0 & 0.67 & 0.68 & 0 & 100 & \textbf{0.62} & \textbf{0.58} & \textbf{100} & 0 & 0 & \textbf{0.62} & \textbf{0.58} & 2 & 98 & 0 \\
    & 18 & 0.81 & 0.80 & 0.81 & 0.80 & 0 & 0.71 & 0.74 & 0 & 100 & \textbf{0.63} & \textbf{0.60} & 0 & 0 & 100 & \textbf{0.63} & \textbf{0.60} & 0 & 100 & 0 \\
    & 20 & 0.82 & 0.81 & 0.82 & 0.81 & 0 & 0.73 & 0.75 & 0 & 100 & \textbf{0.63} & \textbf{0.62} & 0 & 0 & 100 & \textbf{0.63} & \textbf{0.62} & 0 & 54 & 46 \\
    & 30 & 0.85 & 0.86 & 1.00 & 1.00 & 100 & 0.85 & 0.87 & 0 & 100 & \textbf{0.66} & \textbf{0.68} & 0 & 0 & 100 & \textbf{0.66} & \textbf{0.68} & 0 & 0 & 100 \\
    & 40 & 0.88 & 0.87 & 1.00 & 1.00 & 100 & 0.89 & 0.90 & 0 & 100 & \textbf{0.69} & \textbf{0.75} & 0 & 0 & 100 & \textbf{0.69} & \textbf{0.75} & 0 & 0 & 100 \\
    & 50 & 0.89 & 0.91 & 1.00 & 1.00 & 100 & 0.91 & 0.93 & 0 & 100 & \textbf{0.73} & \textbf{0.80} & 0 & 0 & 100 & \textbf{0.73} & \textbf{0.80} & 0 & 0 & 100 \\
    & 60 & 0.90 & 0.92 & 1.00 & 1.00 & 100 & 0.92 & 0.94 & 0 & 100 & \textbf{0.76} & \textbf{0.83} & 0 & 0 & 100 & \textbf{0.76} & \textbf{0.83} & 0 & 1 & 99 \\
    & 70 & 0.91 & 0.92 & 1.00 & 1.00 & 100 & 0.93 & 0.94 & 0 & 100 & \textbf{0.80} & \textbf{0.85} & 0 & 0 & 100 & \textbf{0.80} & \textbf{0.85} & 0 & 0 & 100 \\
    & 80 & 0.92 & 0.93 & 1.00 & 1.00 & 100 & 0.94 & 0.96 & 0 & 100 & \textbf{0.82} & \textbf{0.87} & 0 & 0 & 100 & \textbf{0.82} & \textbf{0.87} & 0 & 0 & 100 \\
    & 90 & 0.93 & 0.94 & 1.00 & 1.00 & 100 & 0.95 & 0.96 & 0 & 100 & \textbf{0.85} & \textbf{0.90} & 0 & 0 & 100 & \textbf{0.85} & \textbf{0.90} & 0 & 0 & 100 \\
    & 100 & 0.94 & 0.94 & 1.00 & 1.00 & 100 & 0.95 & 0.96 & 0 & 100 & \textbf{0.87} & \textbf{0.91} & 0 & 0 & 100 & \textbf{0.87} & \textbf{0.91} & 0 & 0 & 100 \\
    \midrule
    0.8 & 14 & 0.47 & 0.62 & 0.47 & 0.59 & 0 & \textbf{0.32} & 0.44 & 53 & 47 & \textbf{0.32} & \textbf{0.41} & \textbf{100} & 0 & 0 & \textbf{0.32} & \textbf{0.41} & \textbf{100} & 0 & 0 \\
    & 16 & 0.44 & 0.62 & 0.44 & 0.60 & 0 & \textbf{0.29} & 0.42 & 31 & 69 & \textbf{0.29} & \textbf{0.39} & \textbf{100} & 0 & 0 & \textbf{0.29} & \textbf{0.39} & 39 & 61 & 0 \\
    & 18 & 0.45 & 0.63 & 0.44 & 0.61 & 0 & 0.31 & 0.45 & \textbf{6} & 94 & \textbf{0.30} & \textbf{0.40} & 3 & 11 & 86 & \textbf{0.30} & \textbf{0.40} & 0 & 100 & 0 \\
    & 20 & 0.44 & 0.63 & 0.44 & 0.62 & 0 & 0.33 & 0.47 & \textbf{1} & 99 & \textbf{0.31} & \textbf{0.41} & 0 & 0 & 100 & \textbf{0.31} & \textbf{0.41} & 0 & 100 & 0 \\
    & 30 & 0.44 & 0.62 & 1.00 & 1.00 & 100 & 0.40 & 0.58 & 0 & 100 & \textbf{0.32} & \textbf{0.46} & 0 & 0 & 100 & \textbf{0.32} & \textbf{0.46} & 0 & 0 & 100 \\
    & 40 & 0.44 & 0.63 & 1.00 & 1.00 & 100 & 0.44 & 0.65 & 0 & 100 & \textbf{0.33} & \textbf{0.52} & 0 & 0 & 100 & \textbf{0.33} & \textbf{0.52} & 0 & 0 & 100 \\
    & 50 & 0.45 & 0.64 & 1.00 & 1.00 & 100 & 0.46 & 0.69 & 0 & 100 & \textbf{0.35} & \textbf{0.56} & 0 & 0 & 100 & \textbf{0.35} & \textbf{0.56} & 0 & 0 & 100 \\
    & 60 & 0.45 & 0.64 & 1.00 & 1.00 & 100 & 0.46 & 0.71 & 0 & 100 & \textbf{0.35} & \textbf{0.59} & 0 & 0 & 100 & \textbf{0.35} & \textbf{0.59} & 0 & 1 & 99 \\
    & 70 & 0.46 & 0.65 & 1.00 & 1.00 & 100 & 0.47 & 0.71 & 0 & 100 & \textbf{0.36} & \textbf{0.61} & 0 & 0 & 100 & \textbf{0.36} & \textbf{0.61} & 0 & 0 & 100 \\
    & 80 & 0.46 & 0.65 & 1.00 & 1.00 & 100 & 0.47 & 0.71 & 0 & 100 & \textbf{0.37} & \textbf{0.63} & 0 & 0 & 100 & \textbf{0.37} & \textbf{0.63} & 0 & 0 & 100 \\
    & 90 & 0.46 & 0.65 & 1.00 & 1.00 & 100 & 0.46 & 0.72 & 0 & 100 & \textbf{0.37} & \textbf{0.64} & 0 & 0 & 100 & \textbf{0.37} & \textbf{0.64} & 0 & 0 & 100 \\
    & 100 & 0.47 & \textbf{0.66} & 1.00 & 1.00 & 100 & 0.46 & 0.71 & 0 & 100 & \textbf{0.38} & 0.67 & 0 & 0 & 100 & \textbf{0.38} & 0.67 & 0 & 0 & 100 \\
    \bottomrule
    \end{tabular}
    \caption{
        Results on ER graphs, where $|E|$ is the average relative number of edges compared with the original graph, $d_{\rm max}$ is the average relative maximum degree, \#Opt. is the number of optimally solved instances, and \#OM/\#TL is the number of instances where the memory/time limit is reached.
        For $|E|$ and $d_{\rm max}$, the values in the best graph state found by a method are divided by those of the original graph and averaged over all instances with the same parameters.
        Rows for $p = 0.2$ and 50 or more vertices are omitted, since no method is able to find a graph with a smaller number of edges and the maximum degree.
        The best values are in bold.
    }
    \label{tab:er-connoly}
\end{table*}

\begin{table*}[htb]
    \centering
    \small
    \begin{tabular}{lccccccccccccc}
        \toprule
        && \multicolumn{2}{c}{LS} & \multicolumn{2}{c}{LSSF} & \multicolumn{4}{c}{SA-ILP} & \multicolumn{2}{c}{GBFS} & \multicolumn{2}{c}{Orbit} \\
        \cmidrule(lr){3-4} \cmidrule(lr){5-6} \cmidrule(lr){7-10} \cmidrule(lr){11-12} \cmidrule(lr){13-14}
        p & $|V|$ & $|E|$ & $d_{\max}$ & $|E|$ & $d_{\max}$ & $|E|$ & $d_{\max}$ & \#Opt. & \#TL & $|E|$ & $d_{\max}$ & $|E|$ & $d_{\max}$ \\
        \midrule
        0.20 & 13 & 0.94 & 0.90 & 1.31 & 0.89 & \textbf{0.90} & 0.86 & \textbf{100} & 0 & \textbf{0.90} & \textbf{0.82} & \textbf{0.90} & \textbf{0.82} \\
        0.25 & 13 & 0.93 & 0.89 & 1.15 & 0.88 & \textbf{0.87} & 0.83 & \textbf{94} & 6 & \textbf{0.87} & \textbf{0.79} & \textbf{0.87} & \textbf{0.79} \\
        0.30 & 13 & 0.93 & 0.91 & 1.03 & 0.90 & \textbf{0.83} & 0.78 & \textbf{82} & 18 & \textbf{0.83} & \textbf{0.74} & \textbf{0.83} & \textbf{0.74} \\
        0.35 & 13 & 0.91 & 0.87 & 0.95 & 0.87 & 0.80 & 0.73 & \textbf{51} & 49 & \textbf{0.79} & \textbf{0.69} & \textbf{0.79} & \textbf{0.69} \\
        0.40 & 13 & 0.88 & 0.85 & 0.90 & 0.85 & 0.74 & 0.68 & \textbf{36} & 64 & \textbf{0.73} & \textbf{0.65} & \textbf{0.73} & \textbf{0.65} \\
        0.45 & 13 & 0.86 & 0.83 & 0.86 & 0.83 & 0.69 & 0.66 & \textbf{14} & 86 & \textbf{0.68} & \textbf{0.62} & \textbf{0.68} & \textbf{0.62} \\
        0.50 & 13 & 0.79 & 0.78 & 0.80 & 0.78 & 0.64 & 0.62 & \textbf{19} & 81 & \textbf{0.63} & \textbf{0.57} & \textbf{0.63} & \textbf{0.57} \\
        0.55 & 13 & 0.74 & 0.75 & 0.75 & 0.74 & 0.59 & 0.59 & \textbf{11} & 89 & \textbf{0.58} & \textbf{0.55} & \textbf{0.58} & \textbf{0.55} \\
        0.60 & 13 & 0.68 & 0.71 & 0.68 & 0.71 & 0.53 & 0.55 & \textbf{9} & 91 & \textbf{0.52} & \textbf{0.52} & \textbf{0.52} & \textbf{0.52} \\
        0.65 & 13 & 0.63 & 0.66 & 0.62 & 0.66 & 0.48 & 0.51 & \textbf{14} & 86 & \textbf{0.47} & \textbf{0.47} & \textbf{0.47} & \textbf{0.47} \\
        0.70 & 13 & 0.59 & 0.65 & 0.59 & 0.65 & \textbf{0.44} & 0.49 & \textbf{25} & 75 & \textbf{0.44} & \textbf{0.46} & \textbf{0.44} & \textbf{0.46} \\
        0.75 & 13 & 0.50 & 0.63 & 0.50 & 0.60 & \textbf{0.37} & 0.45 & \textbf{54} & 46 & \textbf{0.37} & \textbf{0.43} & \textbf{0.37} & \textbf{0.43} \\
        0.80 & 13 & 0.46 & 0.61 & 0.45 & 0.58 & \textbf{0.31} & 0.43 & \textbf{83} & 17 & \textbf{0.31} & \textbf{0.42} & \textbf{0.31} & \textbf{0.42} \\
        0.85 & 13 & 0.42 & 0.65 & 0.40 & 0.57 & \textbf{0.24} & 0.45 & \textbf{98} & 2 & \textbf{0.24} & \textbf{0.44} & \textbf{0.24} & \textbf{0.44} \\
        0.90 & 13 & 0.38 & 0.72 & 0.32 & 0.68 & \textbf{0.20} & 0.54 & \textbf{100} & 0 & \textbf{0.20} & \textbf{0.53} & \textbf{0.20} & \textbf{0.53} \\
        0.95 & 13 & 0.31 & 0.80 & 0.25 & 0.86 & \textbf{0.17} & \textbf{0.71} & \textbf{100} & 0 & \textbf{0.17} & \textbf{0.71} & \textbf{0.17} & \textbf{0.71} \\
        \bottomrule
        \end{tabular}
        \caption{
       Results on ER graphs with 13 vertices. The definitions of the columns are the same as in Table~\ref{tab:er-connoly}.
       GBFS and Orbit optimally solved all instances. 
    }
    \label{tab:er-sharma}
\end{table*}

\begin{table*}[htb]
    \centering
    \small
    \setlength{\tabcolsep}{1mm}
    \begin{tabular}{lccccccccccccccccccccc}
    \toprule
    && \multicolumn{2}{c}{LS} & \multicolumn{3}{c}{LSSF} & \multicolumn{5}{c}{SA-ILP} & \multicolumn{5}{c}{GBFS} & \multicolumn{4}{c}{Orbit} \\
    \cmidrule(lr){3-4} \cmidrule(lr){5-7} \cmidrule(lr){8-12} \cmidrule(lr){13-17} \cmidrule(lr){18-21}
     & $|V|$ & $|E|$ & $d_{\max}$ & $|E|$ & $d_{\max}$ & \#TL & $|E|$ & $d_{\max}$ & \#Opt. & \#TL & \#OM & $|E|$ & $d_{\max}$ & \#Opt. & \#TL & \#OM & $|E|$ & $d_{\max}$ & \#Opt. & \#TL \\
    \midrule
     BD & 13 & 0.78 & 0.85 & 1.00 & 0.79 & 0 & \textbf{0.68} & 0.72 & 98 & 2 & 0 & \textbf{0.68} & \textbf{0.69} & \textbf{100} & 0 & 0 & \textbf{0.68} & \textbf{0.69} & \textbf{100} & 0 \\
     & 14 & 0.78 & 0.92 & 0.97 & 0.84 & 0 & \textbf{0.69} & 0.79 & 92 & 8 & 0 & \textbf{0.69} & \textbf{0.76} & \textbf{100} & 0 & 0 & \textbf{0.69} & \textbf{0.76} & \textbf{100} & 0 \\
     & 15 & 0.78 & 0.91 & 1.00 & 0.84 & 0 & \textbf{0.69} & 0.81 & 78 & 22 & 0 & \textbf{0.69} & \textbf{0.77} & \textbf{100} & 0 & 0 & \textbf{0.69} & \textbf{0.77} & \textbf{100} & 0 \\
     & 16 & 0.76 & 0.91 & 1.00 & 0.83 & 0 & \textbf{0.68} & 0.80 & 78 & 22 & 0 & \textbf{0.68} & \textbf{0.77} & \textbf{100} & 0 & 0 & \textbf{0.68} & \textbf{0.77} & 98 & 2 \\
     & 17 & 0.79 & 0.91 & 1.04 & 0.84 & 0 & \textbf{0.70} & 0.83 & 74 & 26 & 0 & \textbf{0.70} & \textbf{0.78} & \textbf{99} & 1 & 0 & \textbf{0.70} & \textbf{0.78} & 92 & 8 \\
     & 18 & 0.82 & 0.94 & 1.03 & 0.88 & 0 & \textbf{0.74} & 0.89 & 47 & 53 & 0 & \textbf{0.74} & \textbf{0.83} & \textbf{100} & 0 & 0 & \textbf{0.74} & \textbf{0.83} & 77 & 23 \\
     & 19 & 0.83 & 0.94 & 1.01 & 0.90 & 0 & \textbf{0.75} & 0.92 & 28 & 72 & 0 & \textbf{0.75} & \textbf{0.87} & \textbf{63} & 0 & 37 & \textbf{0.75} & \textbf{0.87} & 39 & 61 \\
     & 20 & 0.83 & 0.98 & 1.01 & 0.90 & 0 & \textbf{0.75} & 0.93 & 21 & 79 & 0 & \textbf{0.75} & \textbf{0.88} & \textbf{39} & 1 & 60 & \textbf{0.75} & \textbf{0.88} & 23 & 77 \\
     & 21 & 0.81 & 0.97 & 1.01 & \textbf{0.89} & 0 & \textbf{0.73} & 0.94 & \textbf{20} & 80 & 0 & \textbf{0.73} & \textbf{0.89} & 19 & 4 & 77 & \textbf{0.73} & \textbf{0.89} & 17 & 83 \\
     & 22 & 0.79 & 0.98 & 1.05 & \textbf{0.86} & 0 & \textbf{0.72} & 0.93 & \textbf{24} & 76 & 0 & \textbf{0.72} & 0.88 & 14 & 0 & 86 & \textbf{0.72} & 0.88 & 9 & 91 \\
     & 23 & 0.80 & 0.97 & 1.04 & 0.92 & 37 & \textbf{0.73} & 0.92 & \textbf{19} & 81 & 0 & \textbf{0.73} & \textbf{0.89} & 4 & 0 & 96 & \textbf{0.73} & \textbf{0.89} & 3 & 97 \\
     & 24 & 0.81 & 1.01 & 1.00 & 1.00 & 100 & \textbf{0.75} & 0.98 & \textbf{7} & 93 & 0 & \textbf{0.75} & \textbf{0.93} & 0 & 0 & 100 & \textbf{0.75} & \textbf{0.93} & 0 & 100 \\
     & 25 & 0.83 & 1.02 & 1.00 & 1.00 & 100 & 0.76 & 0.99 & \textbf{3} & 97 & 0 & \textbf{0.75} & \textbf{0.94} & 1 & 0 & 99 & \textbf{0.75} & \textbf{0.94} & 2 & 98 \\
     \midrule
      DH & 50 & 0.48 & 0.54 & \textbf{0.15} & \textbf{0.20} & 0 & 0.29 & 0.37 & \textbf{21} & 79 & 0 & 0.32 & 0.39 & 0 & 0 & 100 & 0.30 & 0.37 & 0 & 100 \\
     & 100 & 0.42 & 0.48 & \textbf{0.12} & \textbf{0.15} & 0 & 0.24 & 0.31 & 0 & 100 & 0 & 0.31 & 0.39 & 0 & 0 & 100 & 0.29 & 0.36 & 0 & 100 \\
     & 150 & 0.40 & 0.46 & \textbf{0.10} & \textbf{0.12} & 0 & 0.32 & 0.38 & 0 & 100 & 0 & 0.28 & 0.34 & 0 & 0 & 100 & 0.27 & 0.33 & 0 & 100 \\
     & 200 & 0.34 & 0.40 & \textbf{0.08} & \textbf{0.11} & 0 & 0.29 & 0.35 & 0 & 100 & 0 & 0.24 & 0.30 & 0 & 0 & 100 & 0.23 & 0.29 & 0 & 100 \\
     & 250 & 0.34 & 0.41 & \textbf{0.08} & \textbf{0.09} & 0 & 0.32 & 0.37 & 0 & 95 & 5 & 0.23 & 0.31 & 0 & 0 & 100 & 0.23 & 0.30 & 0 & 100 \\
     & 300 & 0.33 & 0.41 & \textbf{0.07} & \textbf{0.09} & 0 & 0.31 & 0.35 & 0 & 71 & 29 & 0.21 & 0.27 & 0 & 0 & 100 & 0.22 & 0.28 & 0 & 100 \\
     & 350 & 0.30 & 0.36 & \textbf{0.07} & \textbf{0.08} & 0 & 0.34 & 0.39 & 0 & 0 & 100 & 0.20 & 0.27 & 0 & 0 & 100 & 0.27 & 0.34 & 0 & 100 \\
     & 400 & 0.32 & 0.38 & \textbf{0.07} & \textbf{0.07} & 0 & 0.37 & 0.40 & 0 & 0 & 100 & 0.21 & 0.28 & 0 & 0 & 100 & 0.37 & 0.45 & 0 & 100 \\
     & 450 & 0.30 & 0.37 & \textbf{0.06} & \textbf{0.07} & 0 & 0.34 & 0.38 & 0 & 6 & 94 & 0.19 & 0.25 & 0 & 0 & 100 & 0.41 & 0.50 & 0 & 100 \\
     & 500 & 0.32 & 0.39 & \textbf{0.07} & \textbf{0.07} & 0 & 0.37 & 0.40 & 0 & 92 & 8 & 0.20 & 0.26 & 0 & 0 & 100 & 0.53 & 0.62 & 0 & 100 \\
    \bottomrule
    \end{tabular}
    \caption{Results on BD and DH graphs. The definitions of the columns are the same as in Table~\ref{tab:er-connoly}. }
    \label{tab:bounded-degree}
\end{table*}

\begin{figure*}[t]
    \centering
    \includegraphics[width=0.79\linewidth]{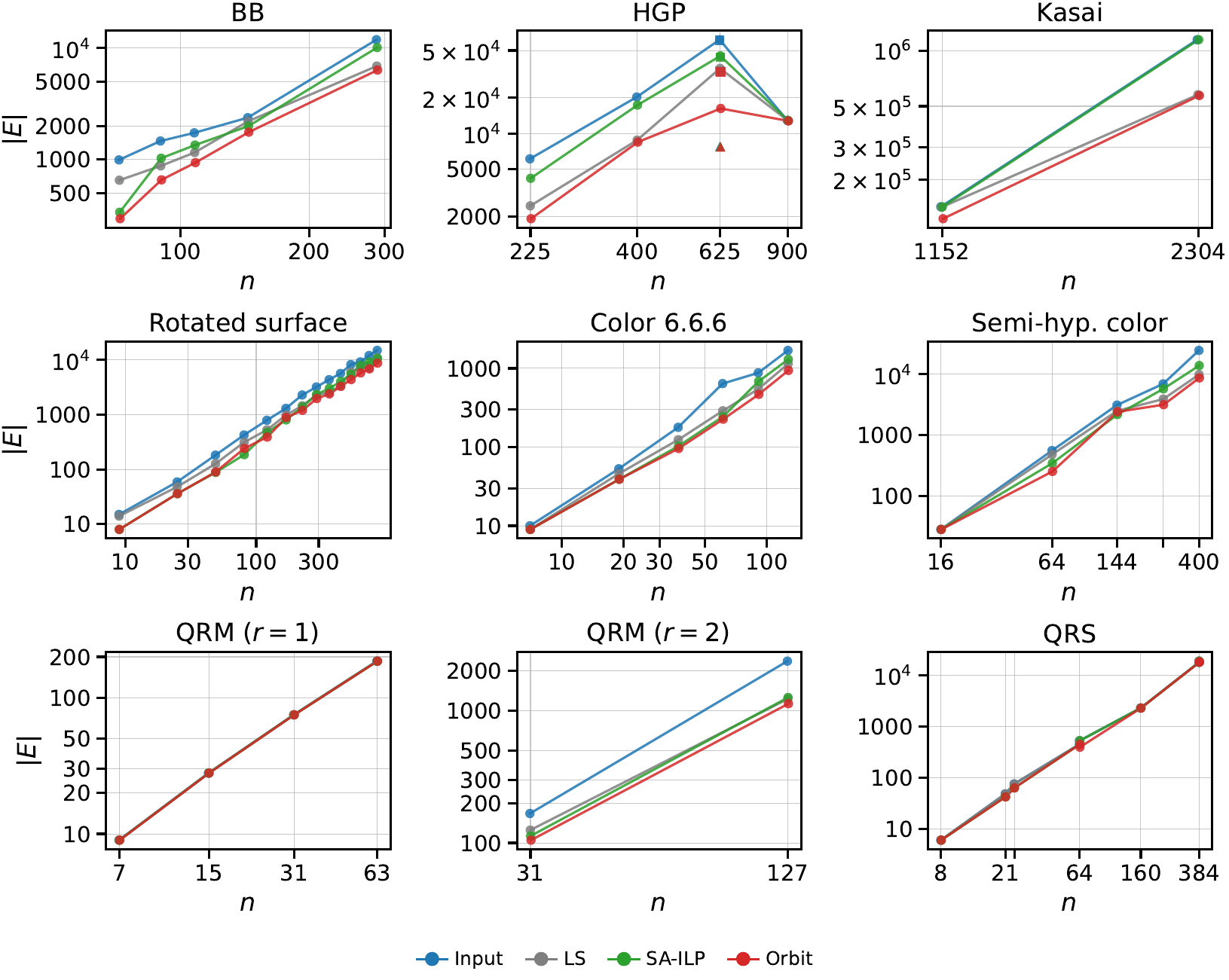}
    \caption{Results on the QECC graphs. The x-axis shows the number of vertices of the input graphs, and the y-axis shows the number of edges of the output graphs. From top to bottom, the rows show non-topological LDPC codes, topological LDPC codes, and algebraic quantum codes.
    Squares and triangles show the HGP $[[625,25,6]]$ and $[[625,25,8]]$ codes from \citeauthor{freire2025optimizing}~\shortcite{freire2025optimizing}, which are not connected by lines.
    GBFS is omitted because its results almost overlap with those of Orbit. LSSF is also omitted because it does not reduce the number of edges in most cases. See Table~\ref{tab:graph-state} for the full results.
    }
    \label{fig:QECC_plot}
\end{figure*}

We experimentally evaluate GBFS and GBFS with orbit search (Orbit).
We implement our approaches in Rust 1.99.1 using jemalloc 0.7.0~\footnote{\url{https://jemalloc.net/}} for the memory allocator.
For the open list, we use a two-level bucketed priority queue, where graph states are prioritized lexicographically by the number of edges and then by the maximum degree. Each bucket contains all graph states with the same number of edges and the maximum degree. Each bucket is maintained as a stack, so ties are broken in the Last-In-First-Out order. The buckets are implemented as a nested array: the outer array is indexed by the number of edges, and each inner array is indexed by the maximum degree.
For graph canonization, we use nauty-Traces-sys 0.11.0,\footnote{\url{https://crates.io/crates/nauty-Traces-sys}} a Rust wrapper of a C library, nauty \cite{McKay2013}.
All experiments are run on a machine with two Intel Xeon Silver 4214R CPUs and Ubuntu 24.04, where each method uses a single thread with a 300-second time limit and an 8 GB memory limit for each instance.

\subsection{Baselines}

As baselines, we evaluate the following methods:
\begin{itemize}
    \item Local search (LS) \cite{connolly2026efficient}. This algorithm enumerates triangles (complete subgraphs with three vertices) in the current graph and performs a local complementation with regard to a vertex in the triangles. If the resulting graph has fewer edges, the current graph is updated.
    We use the authors' Python implementation.\footnote{\raggedright \url{https://github.com/Nicholas-Connolly/Optimizing_complete_multipartite_and_clique_star_graphs}}
    
    \item Local search with split-fuse (LSSF) \cite{connolly2026efficient}. This method decomposes a graph into components called quotient graphs with strong splits, applies LS to each, and then merges them using Type-II fusion operations. It may result in a smaller number of edges and a lower maximum degree than the best values in the LC orbit.
    We use the code in the same repository as LS.
    
    \item ILP to minimize the number of edges using local complementations, warm-starting from the solution found by simulated annealing (SA-ILP) \cite{Sharma2026}. This method guarantees optimality given a sufficient resource limit. 
    We use the authors' Python implementation,\footnote{\url{https://github.com/arr0w-hs/graph_state_optimization}} with Gurobi 13.0.3 \cite{gurobi}.
\end{itemize}
We use Python 3.12.16 to run the baseline methods.

\subsection{Benchmark Instances}

To evaluate the generality of our graph state optimization approach, we consider a variety of graph classes, including random graphs (Erdős–Rényi (ER) graphs), structured graphs, and graphs derived from quantum error-correcting codes. ER graphs have $n$ vertices in total and an edge between each pair of vertices with probability $p$. They were considered as standard benchmarks in previous works~\cite{Sharma2026,connolly2026efficient} and allow us to systematically evaluate performance across different graph sizes and densities. Structured graphs test our approach on specific graph structures to evaluate which graph properties affect the performance of our methods. Graphs derived from quantum error-correcting codes further allow us to assess the effectiveness of our approach on application-relevant instances.
Disconnected graphs are not considered, since it is easily possible to decompose the problem into connected components.

We consider only graphs with $n \geq 13$, since the LC orbits of all graphs with up to 12 vertices have already been enumerated by \citeauthor{adcock2020mapping}~\shortcite{adcock2020mapping}.

For ER graphs, we use three sets of instances, 5200 graphs in total. The first two follow prior work, and the third evaluates scalability to larger $n$.
The first set, based on \citeauthor{Sharma2026}~\shortcite{Sharma2026}, consists of 1600 graphs: 100 graphs with $n=13$ for each $p \in \{ 0.2, 0.25, \ldots, 0.95 \}$.
The second set, based on \citeauthor{connolly2026efficient}~\shortcite{connolly2026efficient}, consists of 1200 graphs: 100 graphs for each combination of $n \in \{ 14, 16, \ldots, 20 \}$ and $p \in \{ 0.2, 0.5, 0.8 \}$.
The third set consists of 2400 graphs: 100 graphs for each combination of $n \in \{ 30, 40,\dots, 100 \}$ and $p \in \{ 0.2, 0.5, 0.8 \}$.
Unlike the original settings, we exclude $p < 0.2$ from the first set, since graphs with small edge probabilities are less likely to be connected. We also exclude $n \leq 12$ from the second set.
We generate all graphs and discard and resample disconnected graphs.

For structured graphs, we use three sets of instances, 3100 graphs in total.
The first set consists of 1300 bounded-degree (BD) graphs: 100 graphs for each $n \in \{13, 14, \dots, 25\}$. The second set consists of 800 graphs locally equivalent to complete multipartite (CM) graphs: 100 graphs for each $n\in\{15, 20, \dots, 50\}$. The last set consists of 1000 distance-hereditary (DH) graphs: 100 graphs for each $n\in\{50, 100, \dots, 500\}$. 
We use the dataset published by \citeauthor{Sharma2026}~\shortcite{Sharma2026} for the BD set,\footnote{\citeauthor{Sharma2026}~\shortcite{Sharma2026} reported results for graphs with up to 16 vertices but published graphs with up to 25 vertices \cite{Sharma2025data}} and generate the CM and DH sets using the code published by \citeauthor{connolly2026efficient}~\shortcite{connolly2026efficient}.

Additionally, we employ practical instances derived from stabilizer states, which are codewords of a family of \emph{quantum error-correcting codes} (QECC) known as stabilizer codes~\cite{gottesman1997stabilizer}. QECC plays a central role in large-scale quantum computing because quantum operations are noisy in general and require error correction during computation, not only communication.

Since every stabilizer state is locally Clifford equivalent to a graph state~\cite{schlingemann2001stabilizer,van2004graphical}, their preparation can be optimized using graph-state optimization techniques~\cite{doherty2026fast}.
We derived corresponding graph states for practical quantum error-correcting codes via Gaussian elimination based on the approach of \citeauthor{doherty2026fast}~\shortcite{doherty2026fast} utilizing Stim package~\cite{gidney2021stim}. 
We generated 48 graph instances using datasets of stabilizer codes in Stim, MQT QECC~\cite{berent2024decoding}, and LDPC~\cite{Roffe_LDPC_Python_tools_2022} packages.

We used instances of quantum low-density parity-check (LDPC) codes and algebraic quantum codes. Quantum LDPC codes have a constant number of stabilizer weight and a constant number of stabilizers per data qubit, even as the code scales. Quantum LDPC codes are promising candidates for fault-tolerant quantum computing because of their potential to reduce the number of physical qubits required for error correction while maintaining a high error-correction threshold~\cite{shor1996fault} under realistic circuit-level noise. The threshold is the maximum physical error rate below which increasing the code size can suppress logical errors. 
In particular, we evaluate \emph{bivariate bicycle} (BB) \emph{codes}~\cite{bravyi2024high}, \emph{hypergraph product} (HGP) \emph{codes}~\cite{kovalev2012improved, tillich2013quantum}\footnote{All HGP instances are constructed as \(\mathrm{HGP}(H,H)\). The regular instances use randomly generated full-row-rank classical LDPC parity-check matrices \(H\) with variable degree 3 and check degree 4. The two additional 625-qubit instances use the published PEG and simulated-annealing (SA) matrices of Ref.~\cite{freire2025optimizing}.}, and \emph{Kasai} codes~\cite{kasai2026breaking}.\footnote{We use the code construction of Ref.~\cite{zhao2026towards} for Kasai codes.}
Among quantum LDPC codes, two-dimensional topological codes~\cite{kitaev2003fault}, such as \emph{rotated surface codes}~\cite{tomita2014low}, \emph{color codes}~\cite{bombin2007exact, bombin2006topological}, and \emph{semi-hyperbolic color codes}~\cite{higgott2026low} are particularly attractive because they can be implemented on quantum processors with nearest-neighbor interactions. Their local structure also provides an interesting setting for evaluating the effectiveness of local search compared with our heuristic search approach.
We also include algebraic quantum codes, such as \emph{quantum Reed–Muller}~\cite{steane1999quantum, hastings2018distillation} and \emph{Reed–Solomon codes}~\cite{grassl1999quantum, grassl2004optimal}, to evaluate our approach on code families with different structural properties where each stabilizer can have high-weight supports depending on the number of physical qubits.

\subsection{Results}

We show the results for the ER graphs in Tables~\ref{tab:er-connoly} and~\ref{tab:er-sharma}, the BD and DH graphs in Table~\ref{tab:bounded-degree}, and the QECC graphs in Figure~\ref{fig:QECC_plot}. We also present the detailed results for the CM and QECC graphs in the appendix (Tables~\ref{tab:cm} and \ref{tab:graph-state}).

With few exceptions, GBFS and Orbit achieve an equal or smaller average number of edges and maximum degree than the baselines. This holds regardless of the values of $p$ and $n$ in the ER graphs (Table~\ref{tab:er-sharma}). 
For ER graphs, BD graphs, and CM graphs, GBFS and Orbit achieve the fewest edges across all parameters. In particular, in ER graphs with $n=13$ (Table~\ref{tab:er-sharma}), GBFS and Orbit optimally solve all graphs. In contrast, the number of instances SA-ILP solves optimally drops from 94 at $p=0.25$ to 9 at $p=0.6$ and recovers to 98 at $p=0.85$. Also, for ER graphs with $p=0.2$ and $n\geq50$, no method reduces the original graphs.

The baselines outperform GBFS in limited cases. For DH graphs, LSSF is a clear winner; indeed, it was designed to efficiently solve DH graphs using a graph decomposition based on a quotient-augmented strong split tree \cite{connolly2026local}. In contrast, on ER graphs, LSSF times out on all instances with $n \geq 30$ while still building the QASST. On BD graphs with 21 to 22 vertices, LSSF also has the best value in the maximum degree while it increases the number of edges on average, possibly due to Type-II fusion operations.
For BD graphs, SA-ILP solves more instances with 21 vertices or more, but GBFS achieves an equal or smaller average number of edges and maximum degree. For ER graphs, baselines achieve better results for specific parameters (e.g., LS achieves the smallest maximum degree for $n=100$ and $p=0.8$).

In our experiments, the ILP phase consumes almost all the run time of SA-ILP. Even on instances that GBFS solves, such as those with $p=0.5$ and $n=14$, SA-ILP times out during the ILP phase in all runs, although it finishes the SA phase in only 0.06 s on average. The SA phase stays short regardless of $n$, taking only 3.77 s on average even for $p=0.5$ and $n=100$.

Comparing GBFS and Orbit, GBFS optimally solves at least as many instances as Orbit in almost all settings. For example, on the ER graphs in Table~\ref{tab:er-connoly}, GBFS solves 662 instances optimally, whereas Orbit solves 434 instances. Nevertheless, Orbit achieves the same number of edges and maximum degree as GBFS in most cases. The exceptions are DH graphs, where Orbit finds fewer edges than GBFS for $n \leq 200$ but more for $n \geq 300$, and some QECC graphs such as color codes, where Orbit finds fewer edges.
The two methods also fail in different ways. For ER graphs, Orbit reaches the time limit on many instances optimally solved by GBFS. GBFS runs out of memory before reaching the time limit on instances where Orbit reaches the time limit with $p=0.2, 0.8$ and 18 to 20 vertices. This is possibly due to increased run time by calculating canonical forms of expanded nodes each time.

For the QECC graphs, Orbit optimally solves the largest number of instances (6 of 48, Table~\ref{tab:graph-state}). LSSF does not reduce the number of edges except for the Color 6.6.6 $[[19,1,5]]$. The results also show tendencies that depend on the code family: GBFS and Orbit achieve the fewest edges and the smallest maximum degrees for non-topological LDPC codes in most cases, while for rotated surface codes, SA-ILP tends to achieve the fewest edges and the smallest maximum degrees on small graphs ($n \leq 169, d\leq 13$). For larger rotated surface codes, GBFS and Orbit outperform others.
For algebraic quantum codes, the original graphs for QRM($r=1$) appear to be already optimal since Orbit solves them optimally without any reduction. SA-ILP, GBFS and Orbit tend to perform similarly on QRS codes. For larger QRS codes, the reduction is small for all methods. 

\subsection{Analysis of Orbit Search}

\begin{figure}[htb]
    \centering
    \includegraphics[width=0.99\linewidth]{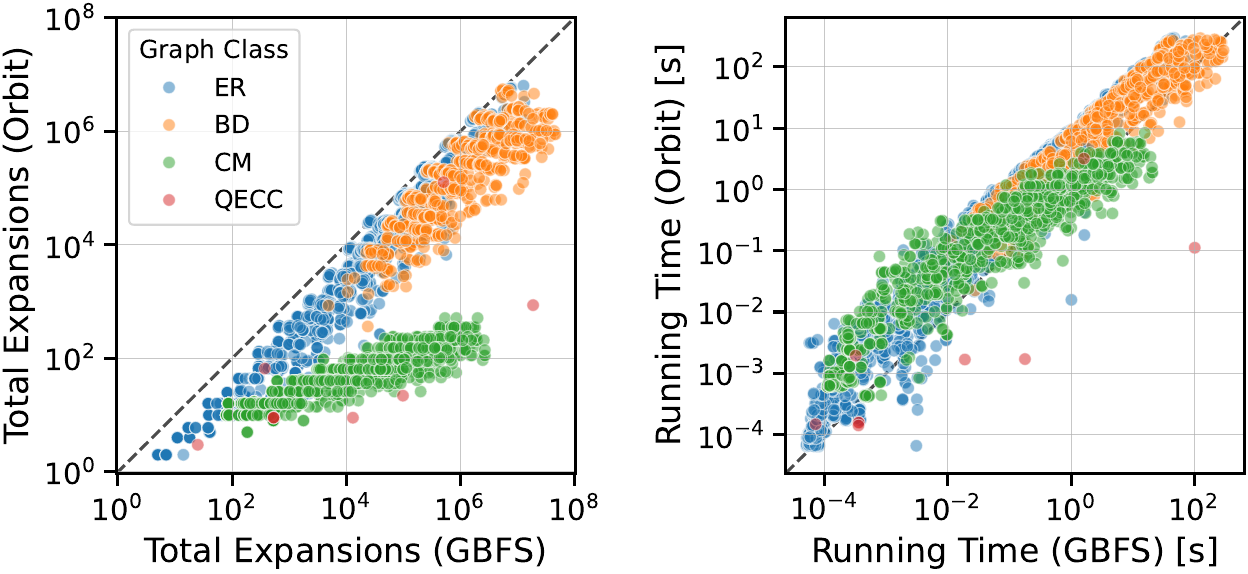}
    \caption{Comparison of the number of expansions and time by GBFS and Orbit to optimally solve each instance.}
    \label{fig:vs-sym}
\end{figure}

We have observed that orbit search improves performance in some DH and QECC graphs while being worse in ER graphs.
We investigate the effect of symmetry breaking by comparing the number of expansions and time to optimally solve an instance in Figure~\ref{fig:vs-sym}.
Using symmetry breaking always reduces the number of expansions, particularly in CM graphs and some BD and QECC graphs.
In terms of time, however, Orbit is not always better, possibly due to runtime overhead introduced by graph canonization.
In particular, in ER graphs, Orbit has higher runtime in most cases.
We conjecture that DH, CM, and QECC graphs tend to have more symmetry than random graphs due to structural restrictions, and thus symmetry breaking is more effective for such instances.

\subsection{Analysis of GBFS's Search Behavior}

\begin{figure}[t]
    \centering
    \includegraphics[width=0.95\linewidth]{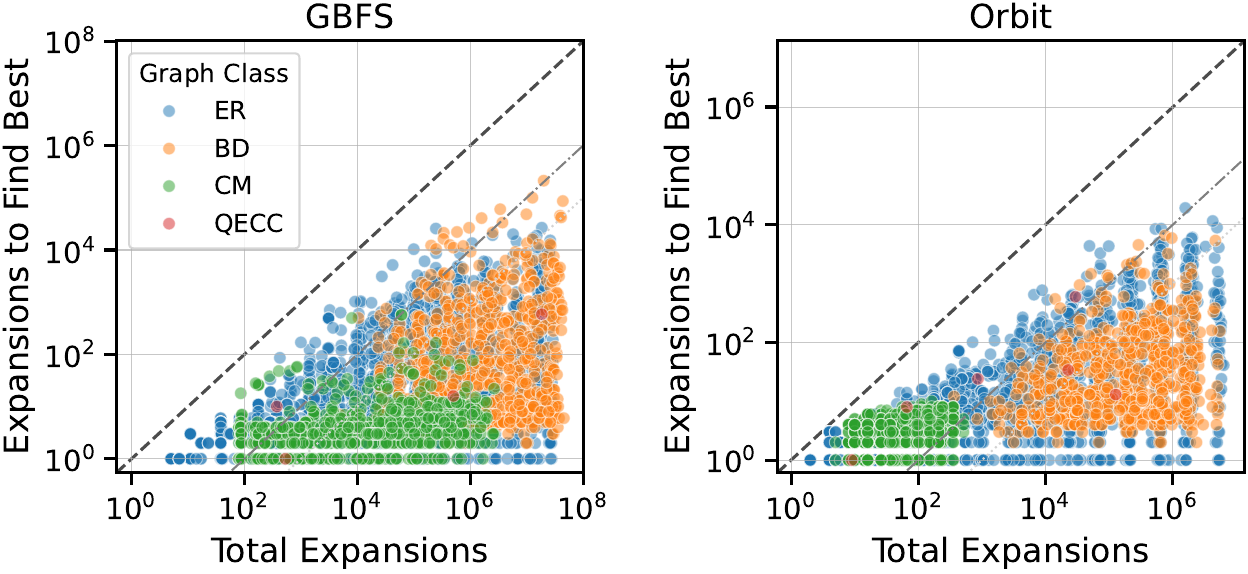}
    \caption{Comparison of the number of expansions to find the first-best state (the first found state with the minimum number of edges) and the number of expansions to optimally solve each instance.}
    \label{fig:find-vs-opt}
\end{figure}

Using optimally solved instances, we compare the number of expansions required to find the first state with the minimum number of edges (we call it the \emph{first-best state}, though it may not minimize the maximum degree) with the number of expansions required to prove optimality by exhausting the open list.
Figure~\ref{fig:find-vs-opt} presents the comparison for both GBFS and Orbit.
In some instances, the first-best state is instantly found with fewer than 10 expansions.
In CM graphs, such instances are the majority, and the tendency is clearer with Orbit.
Even in BD and ER graphs, while some instances require 100 to 10000 expansions, the first-best state is found with more than hundreds of times fewer expansions than proving optimality in most instances.
This result suggests that even when GBFS does not guarantee optimality, it may still find a near-optimal solution within resource limits.

\begin{figure}[t]
    \centering
    \includegraphics[width=0.95\linewidth]{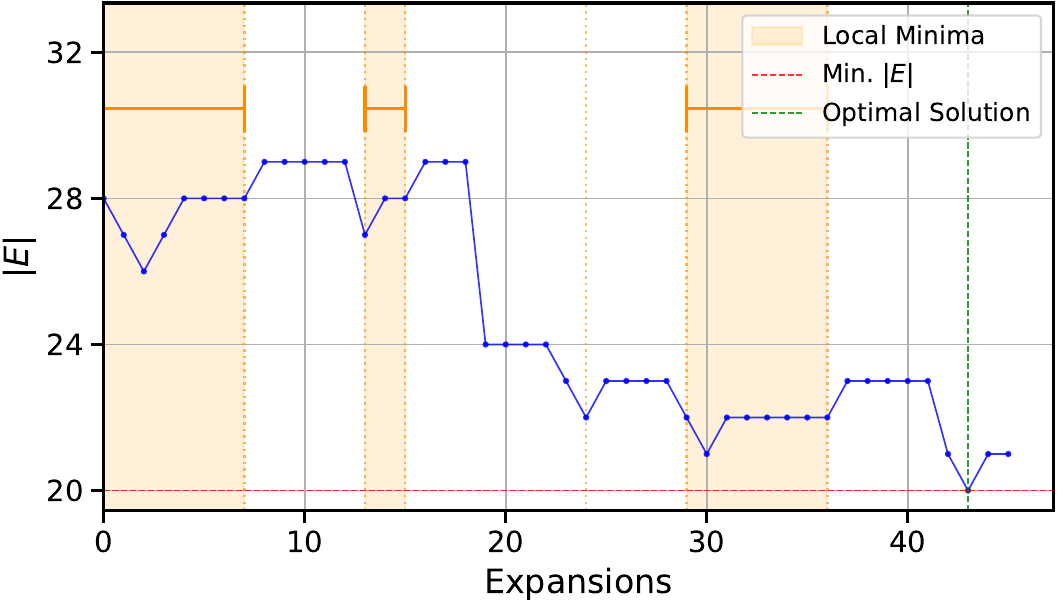}
    \caption{Change of the number of edges in states expanded by GBFS in an optimally solved ER graph instance with 13 vertices and $p=0.2$. }
    \label{fig:score_transitions}
\end{figure}

We further investigate the search behavior by tracing the change in the number of edges in graph states expanded by GBFS.
Figure~\ref{fig:score_transitions} presents such a trace for one particular ER instance with 13 nodes and $p=0.2$, which is optimally solved by GBFS.
The initial graph state has 28 edges, and GBFS reduces it to 26 after two expansions.
However, after that, GBFS needs to explore graph states with more edges (even more than the initial state) to reach a graph state with fewer than 26 edges.
At least in this instance, searching after the first local minimum is necessary to find the first-best state.

To see the general trend beyond a single instance, we count the number of local minima that GBFS explores until finding the first-best state in optimally solved instances.
While there are multiple definitions of local minima  \cite{Wilt2014,Xie2014,Heusner2017,Cohen2018}, for our problem, we use a definition that is intuitive from Figure~\ref{fig:score_transitions} and allows computation from states expanded by GBFS: a sequence of consecutively expanded states forms a local minimum if, before expanding a state with fewer edges than any state in the sequence, GBFS expands a state with more edges than every state in the sequence.
Formally, let $(G_1, \ldots, G_m)$ be the sequence of states expanded by GBFS, where $G_m$ is the first-best state.
Given $i \in \{ 1, \ldots, m - 2 \}$, a state $G_i$ is a \emph{local minimum state} if $|E(G_i)| < |E(G_j)|$ for some $j \in \{ i + 1, \ldots, m - 1 \}$.
A local minimum is a maximal sequence of local minimum states that are expanded consecutively.
In Figure~\ref{fig:score_transitions}, we highlight local minima in orange.

\begin{figure}[t]
    \centering
    \includegraphics[width=0.99\linewidth]{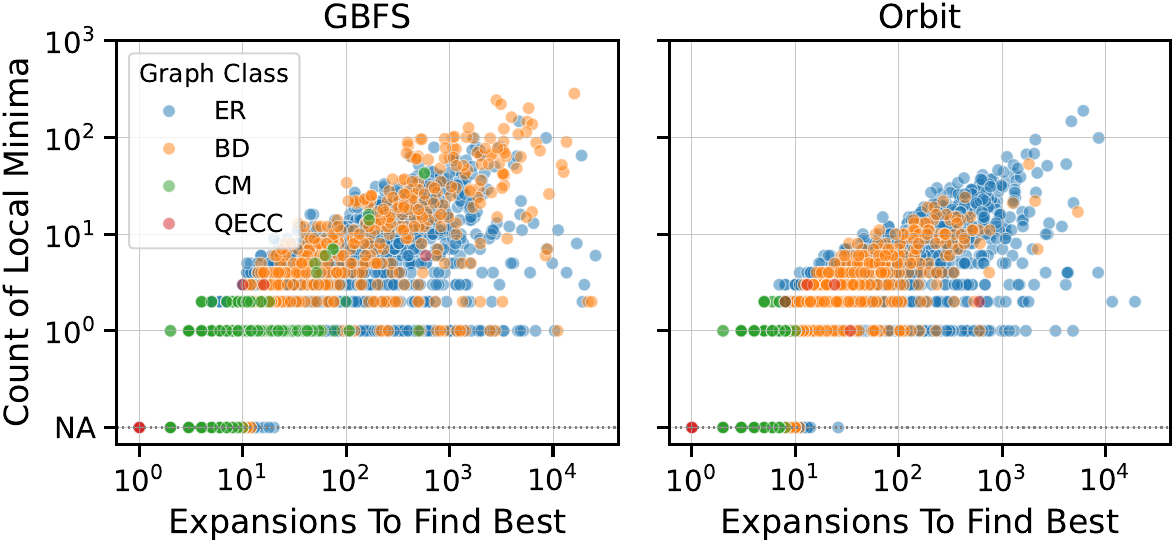}
    \caption{
    Comparison of the number of local minima encountered and the number of expansions to find the first-best state (the first found state with the minimum number of edges).
    Points with `NA' are instances with no local minima.
    }
    \label{fig:lmsize_scatter}
\end{figure}

Figure~\ref{fig:lmsize_scatter} presents the number of local minima explored by GBFS and Orbit in each instance, compared with the number of expansions to find the first-best state.
While no local minima are explored in some instances, a few to hundreds of local minima are explored by GBFS and Orbit in other instances.
This result confirms that stopping at the first local minimum is insufficient to find the first-best state in many instances.
Orbit seems to reduce the number of local minima, particularly in BD graphs.

\section{Conclusion}
We formulated the problem of finding better graph states using a sequence of local complementations as a state space search problem.
We applied greedy best-first search (GBFS) to the problem, demonstrating strong empirical performance compared with existing baselines across multiple graph classes, including those constructed from quantum error-correcting codes.
In addition, in our problem setting, GBFS works as an anytime algorithm, optimally solving instances with up to a few tens of nodes in practice.

In this work, we focused only on local complementations, corresponding to local-Clifford operations in quantum circuits, which are relatively cheap.
Generalizing our approach to optimization of graph states with other types of operators, possibly including more expensive ones, is future work.
Even in such generalized settings, finding better graph states in the LC orbit can be used as preprocessing.
Indeed, \citet{doherty2026fast} used the simulated annealing method \cite{Sharma2026} for this purpose, but we can possibly do better using our GBFS method.
In addition, while \citet{doherty2026fast} tried to minimize the number of expensive operations using heuristic search, their methods do not give a theoretical guarantee of solution quality.
Developing approaches based on optimal, bounded suboptimal, and anytime heuristic search, such as A* \cite{Hart1968}, weighted A* \cite{Pohl1970}, and anytime complete beam search \cite{Zhang1998}, to solve such problems is an interesting direction.

\section*{Acknowledgements}
RK was supported by JST PRESTO, Japan, Grant Number JPMJPR2624.
SN acknowledges support from a JSPS Overseas Research Fellowship.

\bibliography{aaai2027}

\section{Appendices}
We present the results for CM graphs in Table~\ref{tab:cm} and QECC graphs on Table~\ref{tab:graph-state}.

\begin{table*}[tb]
    \centering
    \small
    \begin{tabular}{lccccccccccccc}
    \toprule
    && \multicolumn{2}{c}{LS} & \multicolumn{2}{c}{LSSF} & \multicolumn{4}{c}{SA-ILP} & \multicolumn{2}{c}{GBFS} & \multicolumn{2}{c}{Orbit} \\
    \cmidrule(lr){3-4} \cmidrule(lr){5-6} \cmidrule(lr){7-10} \cmidrule(lr){11-12} \cmidrule(lr){13-14}
     & $|V|$ & $|E|$ & $d_{\max}$ & $|E|$ & $d_{\max}$ & $|E|$ & $d_{\max}$ & \#Opt. & \#TL & $|E|$ & $d_{\max}$ & $|E|$ & $d_{\max}$ \\
    \midrule
     & 15 & 0.64 & 0.84 & \textbf{0.51} & \textbf{0.77} & \textbf{0.51} & 0.79 & \textbf{100} & 0 & \textbf{0.51} & \textbf{0.77} & \textbf{0.51} & \textbf{0.77} \\
     & 20 & 0.54 & 0.77 & \textbf{0.44} & \textbf{0.69} & \textbf{0.44} & \textbf{0.69} & \textbf{100} & 0 & \textbf{0.44} & \textbf{0.69} & \textbf{0.44} & \textbf{0.69} \\
     & 25 & 0.57 & 0.77 & \textbf{0.43} & \textbf{0.69} & \textbf{0.43} & 0.70 & \textbf{100} & 0 & \textbf{0.43} & \textbf{0.69} & \textbf{0.43} & \textbf{0.69} \\
     & 30 & 0.54 & 0.80 & \textbf{0.36} & \textbf{0.69} & \textbf{0.36} & 0.71 & \textbf{99} & 1 & \textbf{0.36} & 0.71 & \textbf{0.36} & 0.71 \\
     & 35 & 0.49 & 0.78 & \textbf{0.38} & \textbf{0.69} & \textbf{0.38} & \textbf{0.69} & \textbf{97} & 3 & \textbf{0.38} & \textbf{0.69} & \textbf{0.38} & \textbf{0.69} \\
     & 40 & 0.48 & 0.77 & \textbf{0.33} & \textbf{0.67} & 0.35 & 0.68 & \textbf{86} & 14 & \textbf{0.33} & 0.68 & \textbf{0.33} & 0.68 \\
     & 45 & 0.50 & 0.80 & \textbf{0.38} & \textbf{0.69} & \textbf{0.38} & 0.70 & \textbf{83} & 17 & \textbf{0.38} & 0.70 & \textbf{0.38} & 0.70 \\
     & 50 & 0.42 & 0.75 & \textbf{0.27} & \textbf{0.65} & 0.28 & 0.66 & \textbf{90} & 10 & \textbf{0.27} & \textbf{0.65} & \textbf{0.27} & \textbf{0.65} \\
    \bottomrule
    \end{tabular}
    \caption{Experimental results on CM graphs. The columns are defined as in Table~\ref{tab:er-connoly}.
    GBFS and Orbit optimally solved all instances. 
    }
    \label{tab:cm}
\end{table*}

\begin{table*}[htb]
    \centering
    \small
    \setlength{\tabcolsep}{1mm}
    \begin{tabular}{lcccccccccccccccc}
    \toprule
    & \multicolumn{3}{c}{Input} & \multicolumn{2}{c}{LS} & \multicolumn{2}{c}{LSSF} & \multicolumn{3}{c}{SA-ILP} & \multicolumn{3}{c}{GBFS} & \multicolumn{3}{c}{Orbit} \\
    \cmidrule(lr){2-4} \cmidrule(lr){5-6} \cmidrule(lr){7-8} \cmidrule(lr){9-11} \cmidrule(lr){12-14} \cmidrule(lr){15-17}
    Code & $|V|$ & $|E|$ & $d_{\max}$ & $|E|$ & $d_{\max}$ & $|E|$ & $d_{\max}$ & $|E|$ & $d_{\max}$ & Stat. & $|E|$ & $d_{\max}$ & Stat. & $|E|$ & $d_{\max}$ & Stat. \\
    \midrule
    BB $[[72,12,6]]$ & 72 & 990 & 45 & 0.66 & 0.87 & 1.00 & 1.00 & 0.34 & \textbf{0.47} & TL & \textbf{0.30} & \textbf{0.47} & OM & \textbf{0.30} & \textbf{0.47} & OM \\
    BB $[[90,8,10]]$ & 90 & 1463 & 49 & 0.60 & 0.88 & 1.00 & 1.00 & 0.70 & 1.00 & TL & \textbf{0.45} & \textbf{0.67} & OM & \textbf{0.45} & \textbf{0.67} & OM \\
    BB $[[108,8,10]]$ & 108 & 1732 & 61 & 0.67 & \textbf{0.57} & 1.00 & 1.00 & 0.78 & 0.90 & TL & \textbf{0.54} & \textbf{0.57} & OM & \textbf{0.54} & \textbf{0.57} & OM \\
    BB $[[144,12,12]]$ & 144 & 2369 & 81 & 0.93 & \textbf{0.88} & 1.00 & 1.00 & 0.84 & 0.90 & TL & \textbf{0.74} & \textbf{0.88} & OM & \textbf{0.74} & \textbf{0.88} & OM \\
    BB $[[288,12,18]]$ & 288 & 11865 & 161 & 0.58 & \textbf{0.73} & 1.00 & 1.00 & 0.85 & 0.95 & OM & \textbf{0.53} & \textbf{0.73} & OM & \textbf{0.53} & \textbf{0.73} & OM \\
    \midrule
    HGP $[[225,9,6]]$ & 225 & 6133 & 108 & 0.40 & 0.59 & 1.00 & 1.00 & 0.69 & 1.01 & OM & \textbf{0.31} & \textbf{0.42} & OM & \textbf{0.31} & \textbf{0.42} & OM \\
    HGP $[[400,16,6]]$ & 400 & 20318 & 196 & 0.43 & 0.74 & 1.00 & 1.00 & 0.86 & 0.98 & OM & \textbf{0.42} & \textbf{0.69} & OM & \textbf{0.42} & \textbf{0.69} & OM \\
    HGP $[[625,25,6]]$ (PEG) & 625 & 61034 & 306 & 0.55 & 0.88 & 1.00 & 1.00 & 0.72 & 1.00 & TL & \textbf{0.54} & \textbf{0.84} & OM & \textbf{0.54} & \textbf{0.84} & OM \\
    HGP $[[625,25,6]]$ (regular) & 625 & 61641 & 323 & 0.57 & 0.89 & 1.00 & 1.00 & 0.73 & 0.93 & TL & \textbf{0.26} & \textbf{0.55} & OM & \textbf{0.26} & \textbf{0.55} & OM \\
    HGP $[[625,25,8]]$ & 625 & 7788 & 48 & 1.00 & \textbf{1.00} & 1.00 & \textbf{1.00} & 1.00 & \textbf{1.00} & TL & \textbf{0.99} & \textbf{1.00} & OM & \textbf{0.99} & \textbf{1.00} & OM \\
    HGP $[[900,36,8]]$ & 900 & 12912 & 52 & 1.00 & \textbf{1.00} & 1.00 & \textbf{1.00} & 1.00 & \textbf{1.00} & TL & \textbf{0.99} & \textbf{1.00} & OM & \textbf{0.99} & \textbf{1.00} & OM \\
    \midrule
    Kasai $[[1152,580,\leq 12]]$ & 1152 & 142562 & 619 & 0.99 & 1.00 & 1.00 & 1.00 & 1.00 & 0.99 & TL & \textbf{0.86} & \textbf{0.76} & OM & \textbf{0.86} & \textbf{0.76} & OM \\
    Kasai $[[2304,1156,\leq 14]]$ & 2304 & 1146252 & 1215 & \textbf{0.50} & 1.01 & 1.00 & 1.00 & 1.00 & 1.00 & TL & \textbf{0.50} & \textbf{0.99} & OM & \textbf{0.50} & \textbf{0.99} & OM \\
    \midrule
    Rotated surface $[[25,1,5]]$ & 25 & 59 & 13 & 0.81 & 0.69 & 1.00 & 1.00 & \textbf{0.61} & 0.54 & TL & \textbf{0.61} & \textbf{0.38} & OM & \textbf{0.61} & \textbf{0.38} & TL \\
    Rotated surface $[[49,1,7]]$ & 49 & 182 & 24 & 0.70 & 0.62 & 1.00 & 1.00 & \textbf{0.48} & \textbf{0.29} & TL & 0.55 & 0.54 & OM & 0.49 & 0.38 & TL \\
    Rotated surface $[[81,1,9]]$ & 81 & 425 & 34 & 0.74 & 0.62 & 1.00 & 1.00 & \textbf{0.44} & \textbf{0.44} & TL & 0.57 & 0.50 & OM & 0.57 & 0.50 & TL \\
    Rotated surface $[[121,1,11]]$ & 121 & 790 & 36 & 0.67 & 0.81 & 1.00 & 1.00 & 0.60 & \textbf{0.47} & TL & 0.50 & 0.53 & OM & \textbf{0.49} & 0.53 & TL \\
    Rotated surface $[[169,1,13]]$ & 169 & 1303 & 59 & 0.75 & 0.53 & 1.00 & 1.00 & \textbf{0.62} & \textbf{0.42} & TL & 0.68 & 0.53 & OM & 0.68 & 0.53 & TL \\
    Rotated surface $[[225,1,15]]$ & 225 & 2287 & 79 & 0.64 & 0.48 & 1.00 & 1.00 & 0.60 & \textbf{0.44} & OM & \textbf{0.53} & 0.47 & OM & \textbf{0.53} & 0.47 & TL \\
    Rotated surface $[[289,1,17]]$ & 289 & 3206 & 71 & 0.71 & 0.69 & 1.00 & 1.00 & 0.73 & 0.69 & OM & 0.63 & \textbf{0.61} & OM & \textbf{0.61} & \textbf{0.61} & TL \\
    Rotated surface $[[361,1,19]]$ & 361 & 4354 & 85 & 0.58 & \textbf{0.48} & 1.00 & 1.00 & 0.69 & 0.62 & OM & \textbf{0.55} & 0.51 & OM & \textbf{0.55} & 0.51 & TL \\
    Rotated surface $[[441,1,21]]$ & 441 & 5656 & 99 & 0.61 & 0.54 & 1.00 & 1.00 & 0.71 & 0.74 & OM & \textbf{0.58} & \textbf{0.49} & OM & \textbf{0.58} & \textbf{0.49} & TL \\
    Rotated surface $[[529,1,23]]$ & 529 & 8224 & 103 & 0.73 & 0.70 & 1.00 & 1.00 & 0.67 & 0.86 & TL & 0.54 & \textbf{0.57} & OM & \textbf{0.53} & \textbf{0.57} & TL \\
    Rotated surface $[[625,1,25]]$ & 625 & 9211 & 136 & 0.80 & 0.61 & 1.00 & 1.00 & 0.86 & 0.75 & TL & \textbf{0.63} & \textbf{0.49} & OM & \textbf{0.63} & \textbf{0.49} & TL \\
    Rotated surface $[[729,1,27]]$ & 729 & 12041 & 130 & 0.60 & \textbf{0.56} & 1.00 & 1.00 & 0.76 & 0.68 & TL & \textbf{0.57} & 0.58 & OM & \textbf{0.57} & 0.58 & TL \\
    Rotated surface $[[841,1,29]]$ & 841 & 14928 & 162 & 0.73 & \textbf{0.56} & 1.00 & 1.00 & 0.72 & 0.61 & TL & \textbf{0.57} & \textbf{0.56} & OM & 0.59 & \textbf{0.56} & TL \\
    \midrule
    Color 6.6.6 $[[19,1,5]]$ & 19 & 53 & 12 & 0.87 & 0.75 & 0.87 & 0.75 & \textbf{0.74} & \textbf{0.50} & TL & \textbf{0.74} & \textbf{0.50} & OM & \textbf{0.74} & \textbf{0.50} & Opt. \\
    Color 6.6.6 $[[37,1,7]]$ & 37 & 178 & 20 & 0.70 & 0.55 & 1.00 & 1.00 & 0.58 & \textbf{0.35} & TL & 0.63 & 0.65 & OM & \textbf{0.54} & \textbf{0.35} & TL \\
    Color 6.6.6 $[[61,1,9]]$ & 61 & 640 & 37 & 0.45 & 0.59 & 1.00 & 1.00 & 0.38 & \textbf{0.41} & TL & 0.37 & 0.46 & OM & \textbf{0.35} & 0.51 & TL \\
    Color 6.6.6 $[[91,1,11]]$ & 91 & 876 & 44 & 0.62 & 0.66 & 1.00 & 1.00 & 0.78 & 1.05 & TL & \textbf{0.53} & \textbf{0.52} & OM & \textbf{0.53} & \textbf{0.52} & TL \\
    Color 6.6.6 $[[127,1,13]]$ & 127 & 1688 & 64 & 0.68 & \textbf{0.67} & 1.00 & 1.00 & 0.77 & 0.88 & TL & 0.62 & 0.73 & OM & \textbf{0.56} & 0.72 & TL \\
    \midrule
    Semi-hyp. color $[[16,8,2]]$ & 16 & 28 & 7 & \textbf{1.00} & \textbf{1.00} & 1.29 & \textbf{1.00} & \textbf{1.00} & \textbf{1.00} & Opt. & \textbf{1.00} & \textbf{1.00} & Opt. & \textbf{1.00} & \textbf{1.00} & Opt. \\
    Semi-hyp. color $[[64,8,6]]$ & 64 & 555 & 41 & 0.86 & 0.85 & 1.00 & 1.00 & 0.62 & 0.56 & TL & \textbf{0.45} & \textbf{0.37} & OM & \textbf{0.45} & \textbf{0.37} & TL \\
    Semi-hyp. color $[[144,8,8]]$ & 144 & 3115 & 81 & 0.80 & \textbf{0.83} & 1.00 & 1.00 & \textbf{0.70} & 0.95 & TL & 0.77 & 0.85 & OM & 0.77 & 0.85 & OM \\
    Semi-hyp. color $[[256,8,12]]$ & 256 & 6959 & 129 & 0.56 & 0.80 & 1.00 & 1.00 & 0.83 & 0.94 & OM & \textbf{0.45} & \textbf{0.71} & OM & \textbf{0.45} & \textbf{0.71} & OM \\
    Semi-hyp. color $[[400,8,?]]$ & 400 & 24712 & 203 & 0.40 & 0.81 & 1.00 & 1.00 & 0.56 & 0.92 & OM & \textbf{0.35} & \textbf{0.70} & OM & \textbf{0.35} & \textbf{0.70} & OM \\
    \midrule
    QRM ($r=1$) $[[15,1,3]]$ & 15 & 28 & 7 & \textbf{1.00} & \textbf{1.00} & \textbf{1.00} & \textbf{1.00} & \textbf{1.00} & \textbf{1.00} & Opt. & \textbf{1.00} & \textbf{1.00} & Opt. & \textbf{1.00} & \textbf{1.00} & Opt. \\
    QRM ($r=1$) $[[31,1,3]]$ & 31 & 75 & 15 & \textbf{1.00} & \textbf{1.00} & \textbf{1.00} & \textbf{1.00} & \textbf{1.00} & \textbf{1.00} & TL & \textbf{1.00} & \textbf{1.00} & OM & \textbf{1.00} & \textbf{1.00} & Opt. \\
    QRM ($r=1$) $[[63,1,3]]$ & 63 & 186 & 31 & \textbf{1.00} & \textbf{1.00} & \textbf{1.00} & \textbf{1.00} & \textbf{1.00} & \textbf{1.00} & TL & \textbf{1.00} & \textbf{1.00} & OM & \textbf{1.00} & \textbf{1.00} & Opt. \\
    \midrule
    QRM ($r=2$) $[[31,1,7]]$ & 31 & 167 & 17 & 0.75 & \textbf{0.65} & 1.00 & 1.00 & 0.68 & \textbf{0.65} & TL & \textbf{0.63} & 0.88 & OM & \textbf{0.63} & 0.88 & TL \\
    QRM ($r=2$) $[[127,1,7]]$ & 127 & 2375 & 73 & 0.52 & 0.88 & 1.00 & 1.00 & 0.53 & 1.00 & TL & \textbf{0.48} & \textbf{0.75} & OM & \textbf{0.48} & \textbf{0.75} & OM \\
    \midrule
    QRS $[[21,9,\geq 3]]$ & 21 & 48 & 9 & 1.00 & 1.00 & 1.00 & 1.00 & \textbf{0.88} & \textbf{0.78} & TL & \textbf{0.88} & \textbf{0.78} & OM & \textbf{0.88} & \textbf{0.78} & TL \\
    QRS $[[24,6,\geq 4]]$ & 24 & 75 & 11 & 1.00 & 1.00 & 1.00 & 1.00 & \textbf{0.84} & \textbf{0.64} & TL & \textbf{0.84} & \textbf{0.64} & OM & \textbf{0.84} & \textbf{0.64} & Opt. \\
    QRS $[[64,8,\geq 8]]$ & 64 & 524 & 23 & 1.00 & 1.00 & 1.00 & 1.00 & 1.00 & 1.00 & TL & \textbf{0.75} & \textbf{0.74} & OM & \textbf{0.75} & \textbf{0.74} & TL \\
    QRS $[[64,32,\geq 5]]$ & 64 & 448 & 29 & \textbf{1.00} & \textbf{1.00} & \textbf{1.00} & \textbf{1.00} & \textbf{1.00} & \textbf{1.00} & TL & \textbf{1.00} & \textbf{1.00} & OM & \textbf{1.00} & \textbf{1.00} & TL \\
    QRS $[[160,80,\geq 9]]$ & 160 & 2280 & 59 & \textbf{1.00} & \textbf{1.00} & \textbf{1.00} & \textbf{1.00} & \textbf{1.00} & \textbf{1.00} & OM & \textbf{1.00} & \textbf{1.00} & OM & \textbf{1.00} & \textbf{1.00} & OM \\
    QRS $[[384,48,\geq 29]]$ & 384 & 18792 & 125 & 1.00 & \textbf{1.00} & 1.00 & \textbf{1.00} & 1.00 & \textbf{1.00} & OM & \textbf{0.97} & \textbf{1.00} & OM & \textbf{0.97} & \textbf{1.00} & OM \\
    QRS $[[384,96,\geq 25]]$ & 384 & 17872 & 131 & \textbf{1.00} & \textbf{1.00} & \textbf{1.00} & \textbf{1.00} & \textbf{1.00} & \textbf{1.00} & OM & \textbf{1.00} & \textbf{1.00} & OM & \textbf{1.00} & \textbf{1.00} & OM \\
    \bottomrule
    \end{tabular}
    \caption{
        Results on QECC graphs.
        `Input' shows the number of vertices ($|V|$), the number of edges ($|E|$), and the maximum degree ($d_{\rm max}$) in the original graph.
        For the evaluated methods, $|E|$, $d_{\rm max}$ are relative to the input as in Table~\ref{tab:er-connoly}. `Stat.' is the solving status, either Opt., TL, or OM, defined as in Table~\ref{tab:er-connoly}.\\   
    }
    \label{tab:graph-state}
\end{table*}
\end{document}